\documentclass[useAMS,usenatbib]{mn2e}

\usepackage{graphicx}

\title[Alignment of Galaxies and Clusters]{Alignment of Galaxies and Clusters}
\author[Hashimoto et al.]{Yasuhiro Hashimoto$^{1,3}$\thanks{Contacting email:
hashimot@saao.ac.za} J. Patrick Henry$^{1,2}$
 and  Hans Boehringer$^{1}$ \\
$^{1}$Max-Planck-Institut f\"ur extraterrestrische Physik,
              Giessenbachstrasse
              D-85748 Garching, Germany\\
$^{2}$Institute for Astronomy, University of Hawaii, 2680 Woodlawn Drive, Honolulu, Hawaii 96822, USA\\
$^{3}$South African Astronomical Observatory, Observatory 7935, South Africa\\
}
\begin{document}

\date{Accepted xx. Received xx}

\pagerange{\pageref{firstpage}--\pageref{lastpage}} \pubyear{2002}

\maketitle

\label{firstpage}

\begin{abstract}
We investigated the
influence of environment on cluster galaxies
by examining the alignment of  
the brightest cluster galaxy (BCG)
position angle with respect
to the host cluster X-ray position angle.
The cluster position angles
were measured using high spatial resolution X-ray data
taken from the Chandra ACIS archive,
that significantly improved the determination of
the cluster shape
compared to the conventional method of using optical
images.
Meanwhile, those of the BCGs were measured
using homogeneous dataset composed of 
high spatial resolution optical images taken with
Suprime-Cam mounted on Subaru 8m telescope.

We found a strong indication of an alignment
between
the cluster X-ray emission and optical light from BCGs,
 while we see no clear direct correlation
 between the degree of ellipticity of X-ray and
 optical BCG morphologies, despite the apparent
 alignment of two elliptical structures.
 We have also investigated
 possible dependence of
 the position angle alignment
 on the X-ray morphology of the clusters,
 and no clear trends are found.
 The fact that no trends are evident
 regarding
 frequency or degree of the alignment with respect to X-ray morphology
 may be consistent with an interpretation
 as a lack of dependence on the dynamical status
 of clusters.
\end{abstract}


\section{Introduction}

It is well established that 
the major axes of galaxy clusters
tend to point toward their nearest neighbour
\citep[e.g.][Hashimoto et al. 2007a]{binggeli1982sao,flin1987agc,rhee1987sea,plionis1994paa,west1995scf,plionis2002csa,chambers2000eca,chambers2002nna}.

\nocite{hashimoto2007icx}

Another alignment effect is that between the
orientation of the brightest cluster galaxy (BCG), or cD galaxy,  and that of 
their parent cluster 
\citep[e.g.][]{sastry1968cas,dressler1978csv,carter1980mcg,binggeli1982sao,struble1990pas,plionis2003gap}.
Similar alignment is also reported for poor groups of galaxies
\citep{fuller1999adg}.  
Numerical work 
\citep[e.g.][]{west1991fsu,vanhaarlem1993vfa,west1994amh,onuora2000acu,splinter1997eao,faltenbacher2005ima}
show that substructure-cluster alignments can occur naturally in
hierarchical clustering models of structure formation
such as the cold dark matter model.

Unfortunately, 
all of these previous galaxy-cluster alignment studies are 
using 
optically-determined cluster position angles,
most of them are  based on the Palomar Observatory Sky Survey (POSS).
Despite the importance of these optical investigations,
individual galaxies may not be the best tracers of the
shape of a cluster. Problems can arise from foreground/background
contamination, as well as the fact that galaxies contribute
discreteness noise.
However, it is believed that the X-ray emitting gas within a cluster
traces its gravitational potential \citep{sarazin1986xec}.
X-ray morphology may then be one of the best observable
phenomenon for determining the cluster shape and orientation.
Indeed, there are several X-ray studies for cluster vs. neighbour-cluster
alignment 
\citep*[e.g.][Hashimoto et al. 2007a]{ulmer1989mar,chambers2000eca,chambers2002nna},
but there are few galaxy-cluster alignment studies
using X-ray morphology, except for 
\citet*{porter1991coa} and \citet*{rhee1991xro}, where they  investigated the BCG-cluster alignment
using low spatial resolution X-ray data, 
as well as traditional cluster shape parameter from 
apparent galaxy distribution, and reported a significant alignment.
Unfortunately, previous X-ray studies
are mostly based on $Einstein$
data. These data are important, but the exposure depths are
small and the spatial resolution is rather low
compared to recently available X-ray data.
The low spatial resolution may not significantly affect relatively
robust measures such as 
position angle 
in a direct way,
but it will critically hinder the accurate
removal of contaminating point sources and
the accurate determination of cluster center, and that
may significantly affect the estimate of 
cluster X-ray morphology including the position angle. 
Hence, a new investigation using
deeper X-ray data with much higher spatial resolution
is needed.

Here we report
a new investigation of galaxy alignment with respect
to its parent cluster, using the cluster position angle and ellipticity
determined by high spatial resolution X-ray data
taken from the Chandra ACIS archive.
Meanwhile, position angle and ellipticity of BCGs are determined from optical
images taken with Subaru 8m telescope.
This paper is organized as follows. In Sec. 2, we describe our main  sample 
and X-ray measures, 
in Sec. 3, details of our optical data are described.
Sec. 4 summarizes our results.
Throughout the paper, we use $H_{o}$ = 70 km s$^{-1}$ Mpc$^{-1}$,
$\Omega_{m}$=0.3, and $\Omega_{\Lambda}$=0.7, unless otherwise stated.

\section[]{The X-ray Sample, X-ray Data Preparation, and X-ray measures}

Here we briefly summarize our main X-ray sample, X-ray data preparation,
and X-ray measures.
\nocite{hashimoto2007rqm}
More detailed descriptions can be found in Hashimoto et al. (2007b).

Almost all clusters are selected from flux-limited X-ray surveys, and X-ray
data are
taken from the Chandra ACIS archive.
A lower limit of z = 0.05 or 0.1 is placed on the redshift to ensure that
a cluster is observed with sufficient field-of-view with ACIS-I or ACIS-S, respectively.
The majority of our sample comes from
the $ROSAT$ Brightest Cluster Sample
\citep[BCS;][]{ebeling1998rbc} and the
Extended $ROSAT$ Brightest Cluster Sample \citep[EBCS;][]{ebeling2000rbc}.
When combined with EBCS, the BCS clusters represent one of the
largest and most complete X-ray selected
cluster samples, which is currently the most frequently
observed by $Chandra$.
To extend our sample to higher redshifts,
additional high-z clusters are selected from various deep  surveys;
10 of these clusters are selected from the
$ROSAT$ Deep Cluster Survey \citep[RDCS;][]{rosati1998rdc},
10 from the $Einstein$ Extended Medium Sensitivity Survey \citep[EMSS;][]{gioia1990eoe,henry1992ems},
14 from the 160 Square Degrees $ROSAT$  Survey \citep{vikhlinin1998cgc},
2 from the Wide Angle $ROSAT$ Pointed Survey \citep[WARPS;][]{perlman2002wsv},
and 1 from  the North Ecliptic Pole survey \citep[NEP;][]{henry2006rne},
RXJ1054 was discovered by \citet{hasinger1998rds},
RXJ1347 was discovered in the $ROSAT$ All Sky Survey \citep{schindler1995das},
and 3C295 has been mapped with $Einstein$ \citet{henry1986xrs}.

The resulting sample contains 120 clusters.
At the final stage of our data processing, to employ our full analysis,
we further applied a selection based on the total counts of cluster emission,
eliminating clusters with very
low signal-to-noise ratio.
Clusters whose center is too close to the edge of the ACISCCD are also removed.
The resulting final sample contains
101 clusters with redshifts between 0.05 - 1.26 (median z = 0.226),
and bolometric luminosity between 1.0 $\times$ 10$^{44}$ -- 1.2 $\times$ 10$^{46}$ erg s$^{-1}$
(median 8.56 $\times$ 10$^{44}$ erg s$^{-1}$).
We reprocessed the level=1 event file retrieved from the archive.
The data were filtered to include only the standard
event grades 0,2,3,4,6 and status 0,
then multiple pointings were merged, if any.
We eliminated time intervals of high background count rate
by performing a 3 $\sigma$ clipping of the
background level.
We corrected the images for exposure variations across the field of view, detector response and telescope vignetting.

We detected point sources using the CIAO routine
celldetect with a signal-to-noise threshold for source detection of three.
 We removed point sources, except for those at the center of the cluster which was
 mostly the peak of the surface brightness distribution rather than a real point source.
The images were then smoothed  with Gaussian $\sigma$=5".
We decided to use isophotal contours to characterize
an object region, instead of a conventional
circular aperture, because we did not want to introduce any
bias in the shape of an object.
To define constant metric scale to all clusters,
we adjusted an extracting threshold in such a way that
the square root of the detected object area times a constant was 0.5 Mpc,
i.e. const$\sqrt{area}$ = 0.5 Mpc.
We chose const =1.5, because
the isophotal limit of a detected object was best represented by
this value.

The morphology of cluster X-ray emission is then characterized objectively by
the position angle, as well as the ellipticity and the asymmetry.
 The position angle is defined by the orientation of the major axis
 measured east from north.
Ellipticity is simply defined by the ratio of semi-major (A) and semi-minor
axis (B) lengths as:
    \begin{eqnarray}
         Elli & = & 1  -B/A
    \end{eqnarray}
   where
   A and B are defined by the
   maximum and minimum spatial {\it rms} of the
   object profile along any direction and computed from the 
   centered-second moments by the formula: \\
    \begin{eqnarray}
         A^2 & = &\frac{\overline{x^2}+\overline{y^2}}{2}+\sqrt{\left(\frac{\overline{x^2}-\overline{y^2}}{2}\right)^2 + \overline{xy}^2}  \\
         B^2 & = &\frac{\overline{x^2}+\overline{y^2}}{2}-\sqrt{\left(\frac{\overline{x^2}-\overline{y^2}}{2}\right)^2 + \overline{xy}^2}
\end{eqnarray}

The asymmetry 
is measured by first rotating a cluster image by 180 degrees around the
object center, then subtracting the rotated image from the original unrotated
one. The residual signals above zero are summed and then normalized.
Please see Hashimoto et al. 2007b for more detailed definitions of 
morphological measures.

\section[]{Optical data}

 To determine the position angle and the ellipticity of BCGs,
 we used optical broad band images taken with
  Supreme-Cam \citep{miyazaki1998cam} on the Subaru telescope.  
  The data were retrieved from Subaru-Mitaka-Okayama-Kiso Archive (SMOKA).
  Reduction software developed by \citet{yagi2002lfn} 
   was used
  for flat-fielding, instrumental distortion correction, differential
  refraction, 
  sky subtraction, and stacking.
The camera covers a 34' $\times$ 27' field of view with a
pixel scale of 0\farcs202.
The photometry is calibrated to Vega system using Landolt standards \citep{landolt1992ups}.
  We refine the original astrometry written as WCS keyword in the
  distributed archival data 
  using using USNO-A2 catalog with positional uncertainties 
  less than  $\sim$ 0.2 arcsec.
The data were taken under various seeing conditions,
and we used only
images with less than $\sim$ 1\farcs2 seeing.
The optical data retrieved from SMOKA contains 30 clusters
with redshifts between 0.08 - 0.9,

Some clusters have observed through many 
wavebands, and that allowed us to investigate
the possible variations of our measures caused by 
waveband shifts.
We have decided to rely primarily on
the R band images  for this alignment study, 
because we found that the  
effect of waveband shift is negligible.

The 
position angle 
and 
ellipticity 
of BCGs are measured exactly the same way as the X-ray cluster emission,
namely,  
the position angle is defined by the orientation of the major axis
 measured east from north, and 
the ellipticity is  defined by  the ratio of semi-major and semi-minor
axis.
Please see Hashimoto et al. 2007b \citep[see also][]{hashimoto1998ies} for further details.
As a precaution,
we investigate the effect of superposed small galaxies
sometimes lying on top of the extended structure of some BCGs,
and we found that
these superposed small galaxies
have little effect on our robust measures such as, position angle 
and ellipticity.

\section[]{Results}

\subsection[]{Systematics}

One of the haunting, yet unfortunately
often lightly treated, problem of any study
comparing complex morphological characteristics of astronomical objects
is the  possible systematics introduced by 
various data quality, exposure times and object redshifts.
Depending on the sensitivity of measures of characteristics,
some susceptible measures  
may be seriously affected by these systematics, producing the misleading
results. 

Unfortunately, investigating the systematics on the
complex characteristics is not an easy task.
To investigate the  systematic effect of, for example, various
exposure times,
one of the  standard approaches is to
simulate an image with a given exposure time
by using an exposure-time-scaled and noise-added model image. 
Unfortunately,
we need to approximate the various characteristics of
a model to the complicated characteristics
of a real object, (and those characteristics are often what we want to 
investigate) and this is
an almost impossible task.

Meanwhile,
if we use the real data, instead of the model,
we will not have this problem.
We can
simulate lower signal-to-noise data caused by a shorter
integration time
by scaling the real data  by the exposure time, and adding Poisson noise
taking each pixel value as the
mean for a Poisson distribution.
However, this simple rescaling and adding-noise process will
produce an image containing
an excessive amount of 
Poisson noise for a given exposure time,
thus lead us to underestimate the data quality.
This inaccurate estimate of noise is 
caused by  the intrinsic noise already presented in the
initial real data.
The intrinsic noise is difficult to be removed
without sacrificing the fine spatial details of
the object.  

Similarly, to investigate the effect of dimming and
smaller angular size caused by higher  redshifts,
in addition to the rest waveband shift effect,
simple rescaling and rebinning
of the real data  will not work, 
because these  manipulations will again produce the
incorrect amount of noise.

Further difficulty associated with simulation
using the real data comes from the fact that 
exposure and redshift effects
are often coupled,
because, in the real observation, low redshift objects are usually observed with
shorter exposures
than high redshift objects.
This coupling
further poses a serious problem,
because
simple standard method of simulating an observation
with `decreased' exposure time will force high redshift
data to get degraded, which greatly reduces signal-to-noise ratio
of already low quality high redshift data.

To circumvent all of these challenging problems,
we developed a very useful simulating technique 
employing
a series of `adaptive scalings'  accompanied by a noise adding
process applied to the real images.
This technique allows us to simulated an image of desired fiducial
exposure time and redshift with correct signal-to-noise ratio
without using a tricky artificial model image,
thus we can easily investigate the effect of various image quality
and/or  easily change the real data to common
fiducial exposure and redshift for easy comparison.
Moreover, the technique can provide us with a powerful
tool for conducting evolutionary studies, enabling us to compare the local 
objects  to the high redshift objects  without degrading 
photon-expensive
high redshift data of low signal-to-noise ratio  
at all.
This method is originally developed for the comparison of
X-ray image data, but can be  used for almost
all kind of imaging data, including optical and NIR images.

Here we briefly describe the method.
Please see Hashimoto et al. 2007b for further details.
To simulate data with integration t=t1, an original unsmoothed image
(including the background) taken with original
integration time t0 was at first rescaled by a factor R$_0$/(1-R$_0$), instead of simple R$_0$,  where R$_0$=t1/t0, t0$>$t1.
That is, an intermediate scaled image I$_1$ was created from the original
unsmoothed image I$_0$ by:
\begin{eqnarray}
  I_1   &=& I_0\frac{R_0}{(1-R_0)}.     
\end{eqnarray}

Poisson noise was then added to this rescaled image by taking each pixel value as the
mean for a Poisson distribution and then randomly selecting a new pixel value from
that distribution. This image was then rescaled again by a factor (1-R$_0$)
to produce an image whose {\it signal} is scaled by R$_0$ relative
to the original image, but
its {\it noise} is approximately scaled by $\sqrt{R_0}$,
assuming that the intrinsic noise initially present in the
real data is Poissonian.

Similarly, to simulate
the dimming effect by the redshift, 
an intermediate scaled image I$_1$
is created from the background subtracted image 
I$_0$ by a pixel-to-pixel manipulation:

\begin{eqnarray}
  I_{1}(x,y) &=& \frac{I_{0}(x,y)^2R_1^2}{[I_{0}(x,y)R_1+B-R_1^2(I_{0}(x,y)+B)]} \\
  where & &\nonumber \\
  R_1&=&[(1+z0)/(1+z1)]^4 
\end{eqnarray}
where z0 and z1 are the original redshift and the  new redshift of the object,
respectively, 
and B is the background.

Finally, to  simulate 
the angular-size change due to the redshift difference
between z0 and z1, the original image will be rebinned 
by a factor R$_2$, then intermediate scaled image will be
created by  rescaling the rebinned image by a factor
1/(R$^2_2$-1), 
before the addition of the  Poisson noise.
For the simulation with `increased' exposure time, 
this factor can be changed to 
R$_3$/(R$_2^2$-R$_3$)
where  R$_3$ = t2/t0, t2$>$t0, where  t2 is the
increased exposure time, and t0 is the original integration time,
and (R$_2^2$ -R$_3$) $>$ 0.
The maximum length of integration time we can `increase' (t2$_{max}$)
is naturally
limited by the original  exposure time and how much
we increase the redshift for the redshift-effect part,
and determined by the relationship,
\begin{eqnarray}
 R_2^2-R_3 = 0, 
\end{eqnarray}
which is equivalent to the case when  no Poisson noise is added after
the rebinning.
Thus,
\begin{eqnarray}
 t2_{max}=t0R_2^2. 
\end{eqnarray}
This t2$_{max}$ can be also used as a rough estimate of
the effective image depth.
The t2$_{max}$ provides an estimate of the image depth
much more effectively  than the conventional simple exposure time
because t2$_{max}$ is related to a quantity
that is affected both by exposure time and redshift,
and thus
enabling us to quantitatively compare exposure times of observations
involving targets at different redshifts (e.g. 100 ksec at z=0.1 and
100 ksec at z=0.9).

\begin{figure}
 \center{
 {\includegraphics[height=7cm,width=5cm,clip,angle=-90]{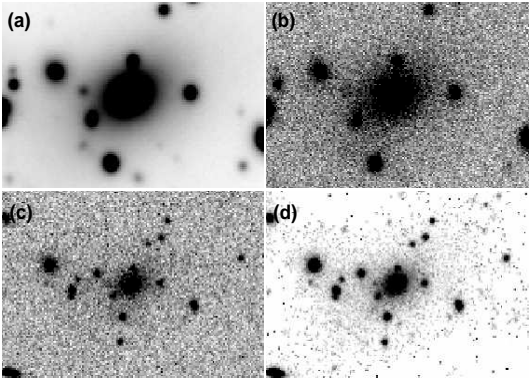}}
 }
 \caption{
Simulating an image with desired exposure time and redshift using the real data:
Even simulating an image with prolonged exposure time is possible
with our adaptive scaling method.
Here, 
optical R band images,
taken with  Subaru Suprime-Cam,
 of the BCG at the center 
of an example cluster (Abell 2219) are shown.
Images with original and modified exposure time and redshift 
are presented with  
north is up and east is left. 
(a) Original image: exptime(t)=240s, and redshift(z)=0.228,
(b) Simulated shorter exposure image with t=10s
(c) Simulated high-z  image with z=0.9, t=240s 
(d) Simulated prolonged exposure at high-z  with t=1092s, z=0.9. 
}
\label{FigTemp}
\end{figure}

\begin{figure}
 \center{
 {\includegraphics[height=7cm,width=5cm,clip,angle=-90]{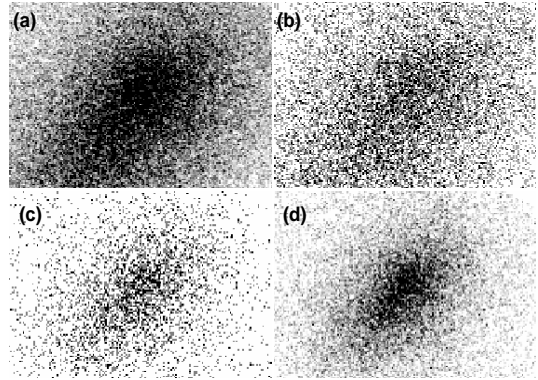}}
 }
 \caption{
Similarly with Fig. 1, 
X-ray images from Chandra ACIS,
of Abell 2219 are shown
with original and modified exposure time and redshift. 
North is up and east is left. 
(a) Original: t=41ks, z=0.228
(b) Simulated shorter exposure: t=10ks (z=0.228)
(c) Simulated High-z image: z=0.9 (t=41ks)
(d) Prolonged exposure at High-z: t=188ks, z=0.9.
}
\label{FigTemp}
\end{figure}

Although we suspected that our ellipticity and position angle
were quite robust, as a precaution 
we investigated  the possible systematics
on these measures
introduced by  various  exposure times and redshifts, using  
our scaling technique described above.

In Fig. 1, 
we demonstrate
our technique of 
simulating desired exposure time and redshift using the real 
optical image of the BCG at the cluster center taken with  Subaru Suprime-Cam.
Original and modified exposure time and redshift  of an example
cluster (Abell 2219) are shown with north  up and east left,
where (a) original image: exptime(t)=240s, and redshift(z)=0.228,
(b) simulated shorter exposure image with t=10s, 
(c) simulated high-z  image with z=0.9, t=240s, and 
(d) simulated prolonged exposure at high-z  with t=1092s, z=0.9.

 Similarly, in Fig. 2, we
 use the real
X-ray images 
 of Abell 2219
from Chandra ACIS,
and simulated various exposures and redshifts, where
(a) original image with t=41ks, z=0.228,
(b) simulated shorter exposure: t=10ks (z=0.228),
(c) simulated High-z image: z=0.9 (t=41ks), and
(d) prolonged exposure at High-z: t=188ks, z=0.9.

Using this technique,
we simulated datasets with various exposure  times
and redshifts, and measured our cluster parameters.
We found that 
our X-ray and optical position angles are 
robust against various exposure times and redshifts.
Similarly, we found that
other morphological measures such as,
the ellipticity and asymmetry are  quite robust, as well.

\subsection[]{Analyses}

\begin{table}
 \tiny
 \caption{Summary of optical cluster sample}
 \label{symbols}
 \begin{tabular}{@{}lcccccc}
  \hline
  Cluster & z &   PA\_X & PA\_BCG & $\Delta$PA$^a$ & Elli\_X & Elli\_BCG \\ 
          & (redshift) &   (degree) & (degree) & (degree)   &          &  \\ 
  \hline
  \hline
 \scriptsize
a2034 & 0.110  &  206.9 & 22.24 & 4.65 & 0.15 & 0.50 \\
a2069 & 0.114  &  327.7 & 331.8 & 4.18 & 0.46 & 0.66 \\
a750 & 0.163  &  249.0 & 249.6 & 0.66 & 0.14 & 0.31 \\
rxj1720 & 0.164  &  355.3 & 32.08 & 36.7 & 0.15 & 0.34 \\
a520 & 0.203  &  192.4 & 228.6 & 36.2 & 0.26 & 0.49 \\
a963 & 0.206  &  175.9 & 347.0 & 8.85 & 0.15 & 0.42 \\
a2261 & 0.224  &  225.8 & 14.48 & 31.3 & 0.14 & 0.13 \\
a2219 & 0.228  &  309.5 & -79.5 & 29.0 & 0.38 & 0.42 \\
a2390 & 0.233  &  298.6 & -57.6 & 3.76 & 0.30 & 0.32 \\
rxj2129 & 0.235  &  246.0 & 65.28 & 0.71 & 0.21 & 0.54 \\
a2631 & 0.278  &  258.2 & 84.27 & 6.07 & 0.29 & 0.41 \\
a1758 & 0.280  &  308.7 & 85.53 & 43.2 & 0.47 & 0.21 \\
a2552 & 0.299  &  201.7 & 216.4 & 14.7 & 0.18 & 0.45 \\
a1722 & 0.327  &  204.4 & 357.7 & 26.6 & 0.27 & 0.14 \\
zwcl3959 & 0.351  &  333.4 & 346.7 & 13.3 & 0.21 & 0.36 \\
a370 & 0.357  &  187.2 & 86.96 & 79.7 & 0.37 & 0.11 \\
rxj1532 & 0.361  &  227.7 & 78.24 & 30.5 & 0.18 & 0.37 \\
zwcl1953 & 0.373  &  351.2 & 306.3 & 44.8 & 0.24 & 0.51 \\
zwcl2661 & 0.382  &  159.5 & 206.9 & 47.3 & 0.12 & 0.43 \\
zwcl0024 & 0.390  &  5.200 & 246.6 & 61.4 & 0.03 & 0.48 \\
rxj2228 & 0.412  &  263.7 & 56.27 & 27.4 & 0.21 & 0.54 \\
rxj1347 & 0.451  &  359.2 & 343.8 & 15.3 & 0.20 & 0.13 \\
ms0451 & 0.540  &  279.9 & 344.9 & 65.0 & 0.26 & 0.04 \\
cl0016 & 0.541  &  228.7 & 244.6 & 15.9 & 0.19 & 0.31 \\
ms2053 & 0.583  &  304.8 & 336.8 & 32.0 & 0.25 & 0.16 \\
rxj1350 & 0.810  &  334.2 & 308.9 & 25.2 & 0.20 & 0.47 \\
rxj1716 & 0.813  &  236.2 & 255.1 & 18.9 & 0.17 & 0.52 \\
ms1054 & 0.830  &  266.1 & 35.96 & 50.1 & 0.40 & 0.51 \\
rxj0152 & 0.835  &  221.6 & 231.0 & 9.46 & 0.58 & 0.29 \\
wga1226 & 0.890  &  284.5 & 264.6 & 19.8 & 0.09 & 0.15 \\
  \hline
  \hline
 \end{tabular}
a: $\Delta$PA is an acute angle of PA\_BCG-PA\_X 

\end{table}

\begin{figure}
 \resizebox{\hsize}{!}{\includegraphics[height=3cm,width=2cm,clip,angle=90]{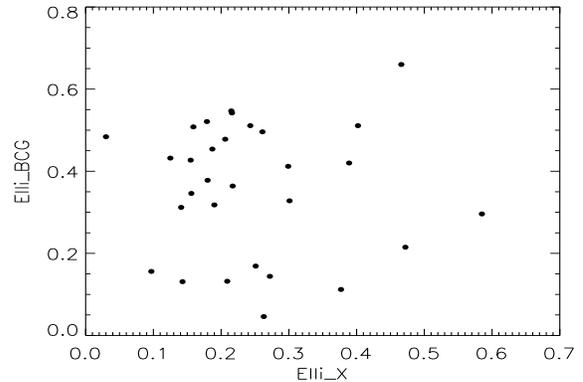}}
 \caption{
Ellipticity of BCGs is plotted against ellipticity of the X-ray morphology
of the host clusters.  
Interestingly, no clear correlation is seen.
}
\label{FigTemp}
\end{figure}

 Table 1 shows a summary of our optical cluster sample,
 where
 $\Delta$PA is an acute relative angle between  
 the position angle of X-ray (PA\_X) and BCG (PA\_BCG),
 namely, the relative position angle differences
 greater than 90 degree are `folded'
 and changed to be acute ranging between 0 and 90 degree. 
 Despite  the robust nature of our measures,
 we modify,
 as a precaution,
 all of the X-ray and optical observations
 to be equivalent to z=0.9 and t=t2$_{max}$
 to eliminate any possible small systematics,
 but otherwise to maximize the image quality.

 In Fig. 3, the ellipticity of cluster X-ray morphology is plotted
 against the ellipticity of optical morphology of BCGs.
 Interestingly, in spite of expected alignment of
two elliptical structures, 
there are a large scatter 
and
we found no strong correlation in  
the relationship between the two ellipticities.

\begin{figure}
 \resizebox{\hsize}{!}{\includegraphics[height=3cm,width=2cm,clip,angle=90]{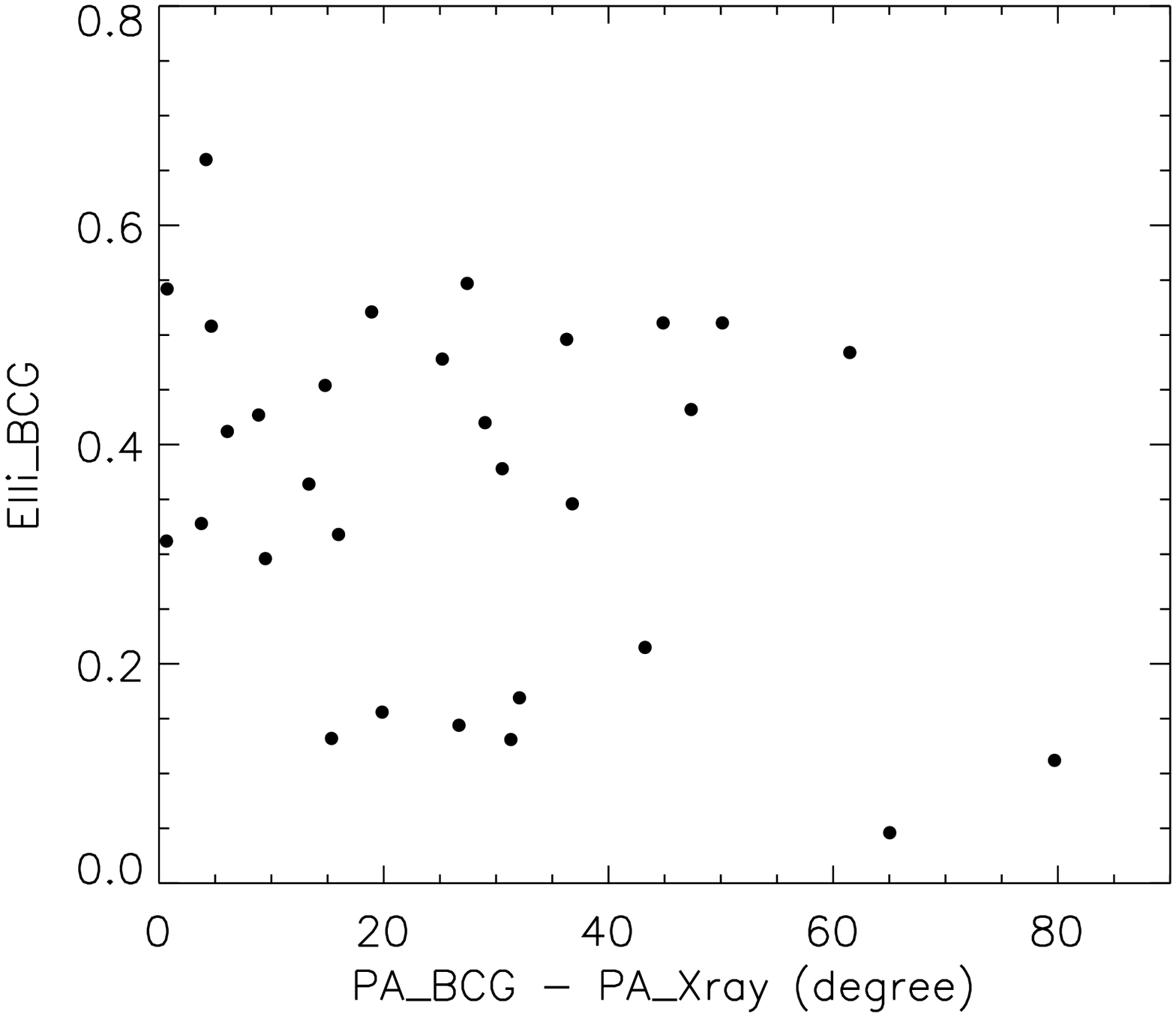}}
 \resizebox{\hsize}{!}{\includegraphics[height=3cm,width=2cm,clip,angle=90]{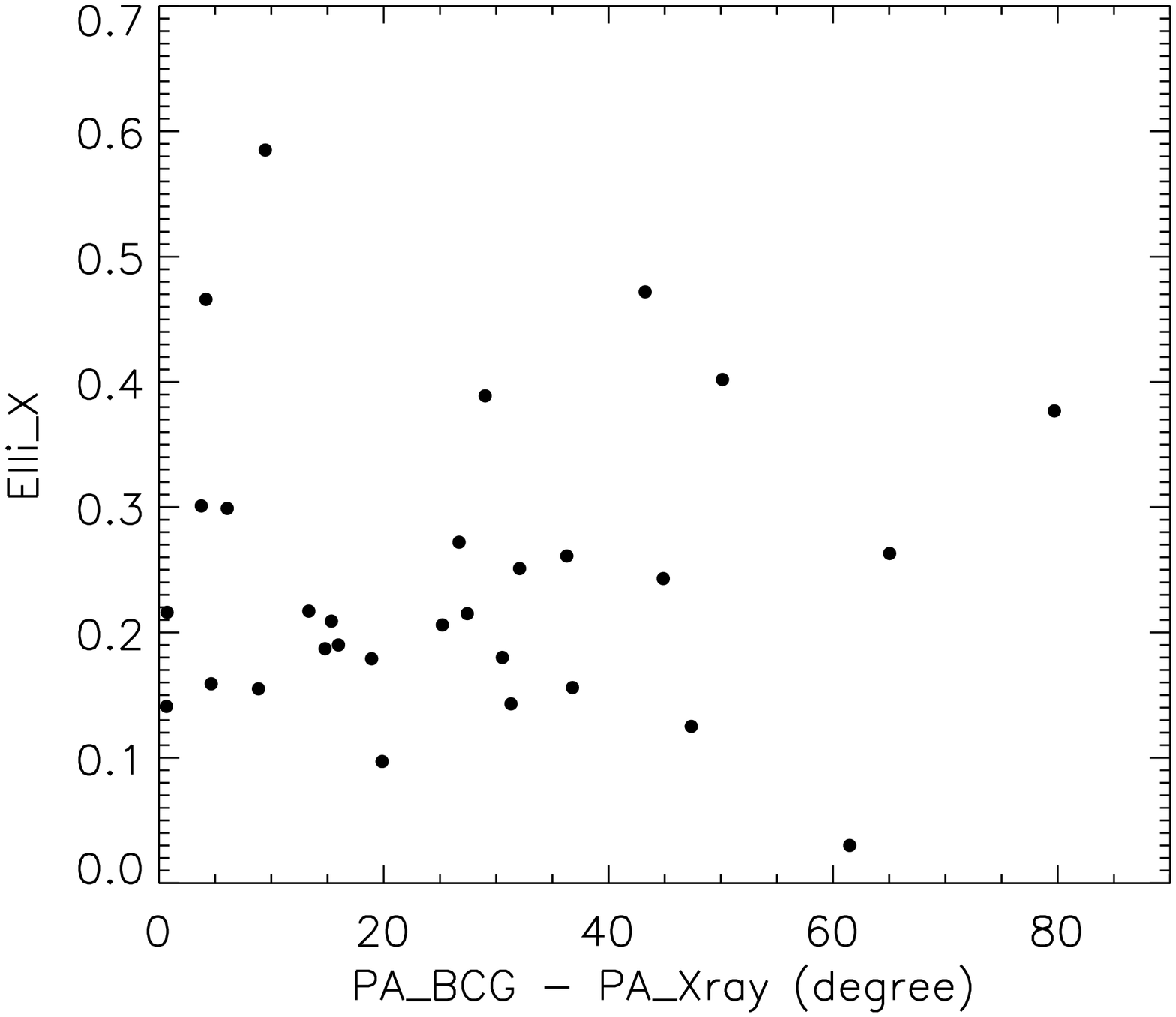}}
 \caption{
The acute position angle difference plotted
 versus ellipticity of BCGs (top panel)
and ellipticity of cluster X-ray emission. 
 The position angle difference is determined
by the difference between  the cluster X-ray position angle
and the position angle of BCG galaxy.
Figures illustrate that 
clusters with the position angle difference
less than 45 degree tend to be more abundant,
particularly for high ellipticity BCGs or clusters,
implying
that cluster X-ray emission  and optical light from BCG are
aligned.
}
\label{FigTemp}
\end{figure}

\begin{figure}
 \resizebox{\hsize}{!}{\includegraphics[height=2cm,width=3cm,clip,angle=0]{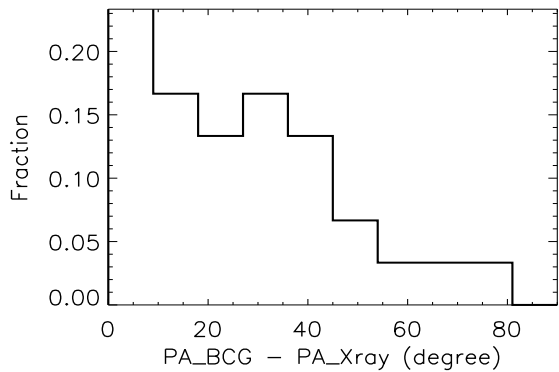}}
 \caption{
The frequency distribution of the  position angle difference.
 There is a tendency that
 we  
 have more clusters with an angle difference less than 45 degrees,
 consistent with the observation made in Fig. 4 implying
that cluster X-ray emission  and optical light from BCG are
aligned.
}
\label{FigTemp}
\end{figure}
 
 Fig. 4 shows the acute relative position angle difference 
 between cluster X-ray morphology and BCG morphology
 plotted
 against  the ellipticity of BCGs (top panel) and the ellipticity of 
cluster X-ray morphology
 (bottom panel).
 Fig. 4 shows that the position angle difference
tends to be  smaller than 45 degree, implying
that BCGs tend to elongated in the same direction of the X-ray
distribution of  their host clusters,
particularly for clusters exhibiting relatively high ellipticity 
in their optical
BCG morphology and/or in their cluster X-ray  morphology.
Meanwhile,
for clusters with very low BCG or X-ray ellipticity (ellipticity $<$ 0.1) 
position angle are generally poorly determined,
and thus position angle difference can be inaccurate.

 Fig. 5 shows the frequency distribution of the  position angle difference.
 There is a strong tendency that we 
 have more clusters with an angle difference less than 45 degrees,
 consistent with the observation made in Fig. 4. 

 To test this trend more rigorously, we first employed
 the Kolmogorov-Smirnov (K-S) test.
 The null hypothesis here is that our sample can be drawn from
a parent population of random position angle differences.
 However,
 the K-S test detects the  deviation from the parent
 population (here the population of  random position angle differences),
 thus it may loose some sensitivity for testing the cluster alignment,
 where it is likely 
 that position angle difference is systematically lower than the
 random sample.
To increase the sensitivity to an alignment signal,
 as a second test we employed the Wilcoxon-Mann-Whitney  rank-sum test.
The null hypothesis of this test is that the position angle difference
is not systematically
smaller or larger than the random sample. Therefore the test is insensitive
to an excess of angles around the mean (i.e. 45 deg).
 When applied to our sample,
 both K-S and rank-sum tests show, not surprisingly  the strong 
 alignment signals,
 and we find that the null hypothesis can be rejected with 99.93\% and
 99.99\% confidence, respectively, 
thus confirming that BCGs are significantly aligned to the X-ray
 emissions of the host clusters.
 We have also investigated the alignment of other luminous non-BCG galaxies 
 to the X-ray emissions and we found no significant alignment.
 The results are summarized in table 2, where LG2
 is the second brightest galaxy, LG3 is the third brightest galaxy,
 and LGn is the n-th brightest
 galaxy  within a projected distance of 1 Mpc from the X-ray center.

\begin{table}
 \caption{Significance levels of the alignment for various luminous galaxies } 
 \label{symbols}
 \begin{tabular}{@{}lccccc}
  \hline
  Statistics & BCG & LG2 &  LG3 & LG4 & LG5  \\ 
  \hline
  \hline
  K-S      & 99.93  & 81.67 & 49.93 & 4.22 & 8.99 \\
  Rank Sum & 99.99  & 83.06 & 57.09 & 54.51 & 63.33 \\
  \hline
  \hline
 \end{tabular}
\end{table}

\begin{figure}
 \resizebox{\hsize}{!}{\includegraphics[height=3cm,width=2cm,clip,angle=90]{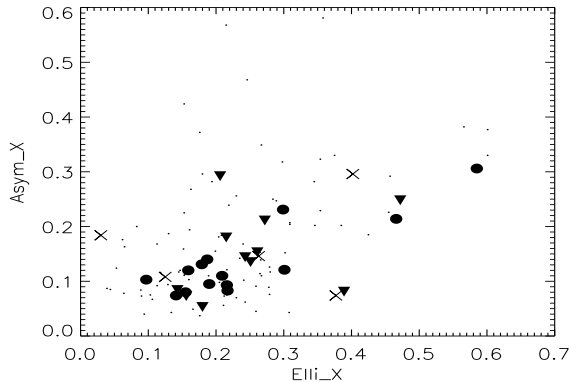}}
 \caption{
 X-ray morphology versus BCG alignment:
 Cluster X-ray asymmetry is plotted against cluster X-ray ellipticity.
 Large solid circles are clusters showing strong alignment with
 BCG-to-cluster position angle difference less than 20 degree, while
 solid triangles are clusters showing the ``modest''alignment, but
with position angle difference between 20 and 45 degree.
 Crosses represent clusters with no sign of the alignment, 
 and small dots are clusters in our main sample without optical Subaru data.
 No clear trends are visible 
 regarding 
 frequency or degree of the alignment with respect to X-ray morphology,
 which can be interpreted as lack of dependence on the dynamical status
 of clusters. 
}
\label{FigTemp}
\end{figure}

 In Fig. 6, we investigated 
 possible dependence of 
 the position angle alignment
 on the X-ray morphology of the clusters.
 In Fig. 6,
 the cluster X-ray asymmetry is plotted against cluster X-ray ellipticity.
 Large solid circles are clusters showing strong alignment 
 between the cluster and BCG
 with
 position angle difference less than 20 degree, while
 solid triangles are clusters showing the alignment, but
with position angle difference between 20 and 45 degree.
 Crosses represent clusters with no sign of the alignment,
 and small dots are clusters in our main X-ray sample without optical Subaru data.
 No clear trends are evident 
 regarding
 frequency or degree of the alignment with respect to X-ray morphology.

 We have also attempted to investigate 
 possible dependence of the alignment
 on cluster redshifts.
We found that
 for clusters less than  z=0.35, 
 both  
 K-S and rank-sum tests show 
 that the null hypothesis can be rejected with 99.92\% and
 99.99\% confidence, respectively,
 while for clusters greater than or equal to  z=0.35, 
 alignment signals are somewhat weaker
 that the null hypothesis can be rejected with 93.88\% and
 83.69\% confidence, respectively for K-S and rank-sum tests.
 Similarly, we have investigated the dependence
 of the alignment on the cluster X-ray bolometric luminosity (Lbol),
 and we did not find any significant trend:
 for clusters with  Lbol greater than or equal to 2 $\times$10$^{45}$ erg/s, 
 the null hypothesis can be rejected with 93.45\% and
 96.42\% confidence, 
 while for 
 clusters with Lbol smaller than 2 $\times$10$^{45}$ erg/s,
 the null hypothesis can be rejected with 99.78\% and
 99.90\% confidence, 
 respectively for K-S and rank-sum tests.

\section{Summary}
We investigated the
influence of environment on cluster galaxies  
by examining the alignment of the  BCG position angle with respect
to the host cluster X-ray position angle.
The cluster position angles 
were measured using high spatial resolution X-ray data
taken from the Chandra ACIS archive,
that significantly improved the determination of
the cluster shape  
compared to the conventional method of using optical
images.
Meanwhile, those of the BCGs were measured
using high spatial resolution optical images taken with
Suprime-Cam mounted on Subaru 8m telescope.

We found a strong indication of an alignment 
between 
the cluster X-ray emission and optical light from BCGs,
 while we see no clear direct correlation
 between the ellipticity of X-ray morphology and
 optical BCG morphology despite of the apparent 
 alignment of two elliptical structures.
In the hierarchical structure formation models,
the alignment effect could be produced  by 
clustering models of structure formation 
such as the cold dark matter model
\citep[e.g.][]{salvadorsole1993tie,west1994amh,usami1997ter,onuora2000acu,faltenbacher2002cog,faltenbacher2005ima}.
The existence of the alignment effects
is also consistent with a cosmic structure formation model such
as the hot dark matter model \citep[e.g.][]{zeldovich1970gia},  where
clusters and galaxies form by fragmentation in already flattened sheet-
and filament-like structures.

 We have also investigated
 possible dependence of
 the position angle alignment
 on the X-ray morphology of the clusters,
 and no clear trends are found.
 If the X-ray morphology of clusters 
 reflects dynamical status of clusters 
\citep[e.g.][]{hashimoto2004aca},
 the fact that no trends are evident
 regarding
 frequency or degree of the alignment with respect to X-ray morphology
 may be consistent with an interpretation
 as a lack of dependence of alignment on the dynamical status
 of clusters.
 Primordial galaxy alignments in clusters can be damped by various
 mechanisms such as  the 
 exchange of angular momentum in galaxy encounters, violent relaxation,
 and secondary infall \citep[e.g.][]{quinn1992gaa,coutts1996sad}  over a Hubble time.
 Thus, in highly relaxed clusters, 
  we might naively expect to observe weaker primordial galaxy alignment,
  because there has been sufficient time to mix the phases.
  The fact that we do not see any significant alignment trend 
  with respect to the X-ray morphology may provide an important
   constrain on these damping scenarios.

\section*{Acknowledgments}
 This work is based in part on data collected at Subaru Telescope and obtained from the SMOKA, which is operated by the Astronomy Data Center, National Astronomical Observatory of Japan.
 We thank our referee, Dr. Michael West for his comments, which improved the
  manuscript.
 YH thanks Hiromi Hashimoto for the help retrieving the data from  SMOKA.


\end{document}